# A Case for Peering of Content Delivery Networks

Rajkumar Buyya[1], Al-Mukaddim Khan Pathan[1], James Broberg[2] and Zahir Tari[2]

[1]**Gri**d Computing and **D**istributed **S**ystems (GRIDS) Laboratory
Department of Computer Science and Software Engineering
The University of Melbourne, Carlton, VIC 3053, Australia
{raj, apathan}@csse.unimelb.edu.au

[2]Distributed Systems and Networking Discipline
School of Computer Science and Information Technology
RMIT University, VIC 3083, Australia
{jbroberg, zahirt}@cs.rmit.edu.au

## 1. Insight into Content Delivery Networks

Content Delivery Networks (CDN) [9][10], which evolved first in 1998 [3], replicate content over several mirrored Web servers (i.e., surrogate servers) strategically placed at various locations in order to deal with *flash crowds* [1]. Geographically distributing the web servers' facilities is a method commonly used by service providers to improve performance and scalability. A CDN has some combination of a content-delivery infrastructure, a request-routing infrastructure, a distribution infrastructure and an accounting infrastructure. CDNs improve network performance by maximizing bandwidth, improving accessibility and maintaining correctness through content replication. Thus CDNs offer fast and reliable applications and services by distributing content to cache servers located close to end-users [4].

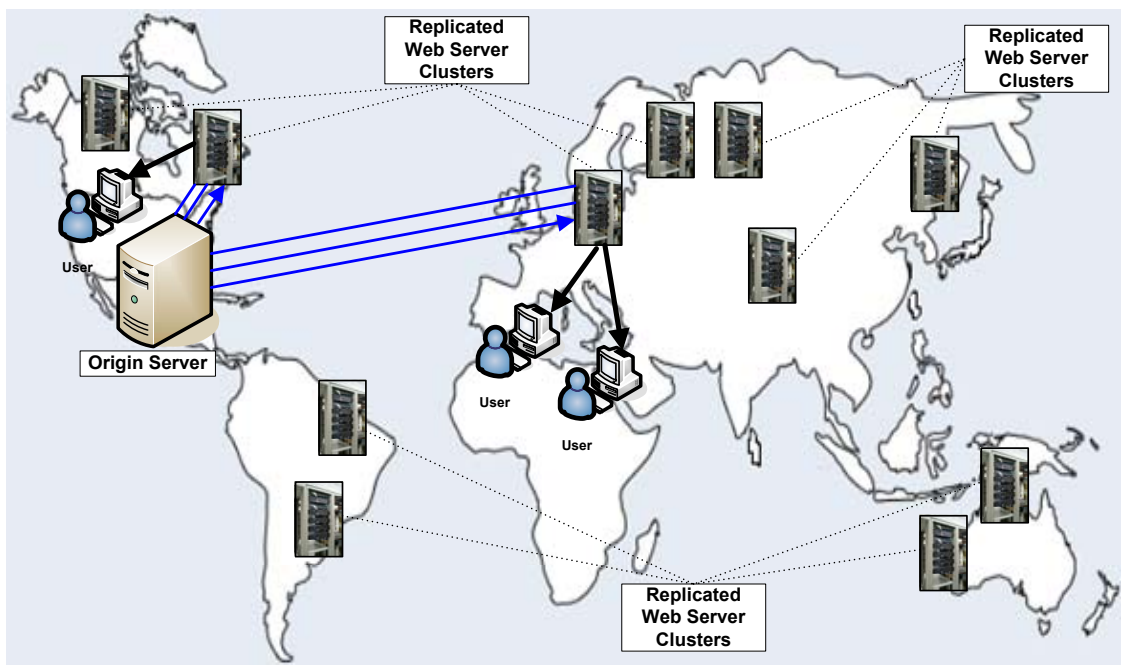

**Figure 1:** Abstract architecture of a Content Delivery Network (CDN)

Figure 1 shows a typical content delivery environment where the replicated Web server clusters are located at the edge of the network to which the end-users are connected. In such CDN environment, Web content based on user requests are fetched from the origin server and a user is served with the content from the nearby replicated Web server. Thus the users end up communicating with a replicated CDN server close to them and retrieves files from that server.

## 2. Motivations for Peering CDNs

Existing Content Delivery Networks (CDN) are proprietary in nature. They are owned and operated by individual companies. They have created their own closed delivery network, which is expensive to setup and maintain. Although there are many commercial CDN providers, they do not cooperate in delivering content to end-users in a scalable manner. In addition, content providers are typically subscribed to one of the CDN providers and are unable to utilize services of multiple CDN providers at the same time. Such a closed, non-cooperative model results in creation of "islands" of CDNs. Running a global CDN requires enormous amount of capital and labor. To compromise expense, CDN providers should partner together so that each can supply and receive services in a cooperative and collaborative manner that one CDN cannot provide to content providers otherwise.

Commercial CDNs charge customers for their services, and in turn they are bound with strong commitment with their end-users to meet the negotiated Service Level Agreement (SLA). An SLA is a part of contract between the service providers and their consumers. It describes provider's commitment and specifies penalties if those commitments are not met. The objective of a CDN is to satisfy its customers with competitive services. If a particular CDN provider is unable to provide quality service to the client requests, it may result in SLA violation and adverse business impact. In such scenarios, one CDN provider should peer with other CDN provider(s) which has caching servers located near to the client and serve the client's request, meeting the Quality of Service (QoS) requirements.

CDN internetworking [8][11] can better be described by illustrating a scenario depicted in Figure 2. Consider that the ICC Cricket World Cup 2007 is to be held in the Caribbean and www.cricinfo.com is supposed to provide live media coverage of the cricket matches from there. As a content provider www.cricinfo.com has exclusive service level agreement with the CDN provider, Akamai [6][7]. However, Akamai does not have any Point of Presence (POP) in Trinidad and Tobago (one of the Caribbean islands) where most of the cricket matches would be held. Akamai management may decide to place its surrogates in Trinidad and Tobago or they may use their edge servers which are in other Caribbean island (e.g. St. Lucia) in order to serve the users in Jamaica. In the first case, placement of new surrogates only due to a particular event would cost much for the CDN provider, which may be redundant after the event. On the other hand, Akamai might be at risk of losing reputation due to the inability to provide quality service according to the client requests; which may result in SLA violation and adverse business impact. If another CDN provider, Mirror-Image has its POP in Trinidad and Tobago, Akamai may partner together with Mirror-Image's edge servers in order to provide quality services based on negotiated SLA. Thus, through collaboration with another CDN provider, content networks can emphasize on customer satisfaction meeting their QoS requirements, and hence it can minimize the exposed business impact of service level violations.

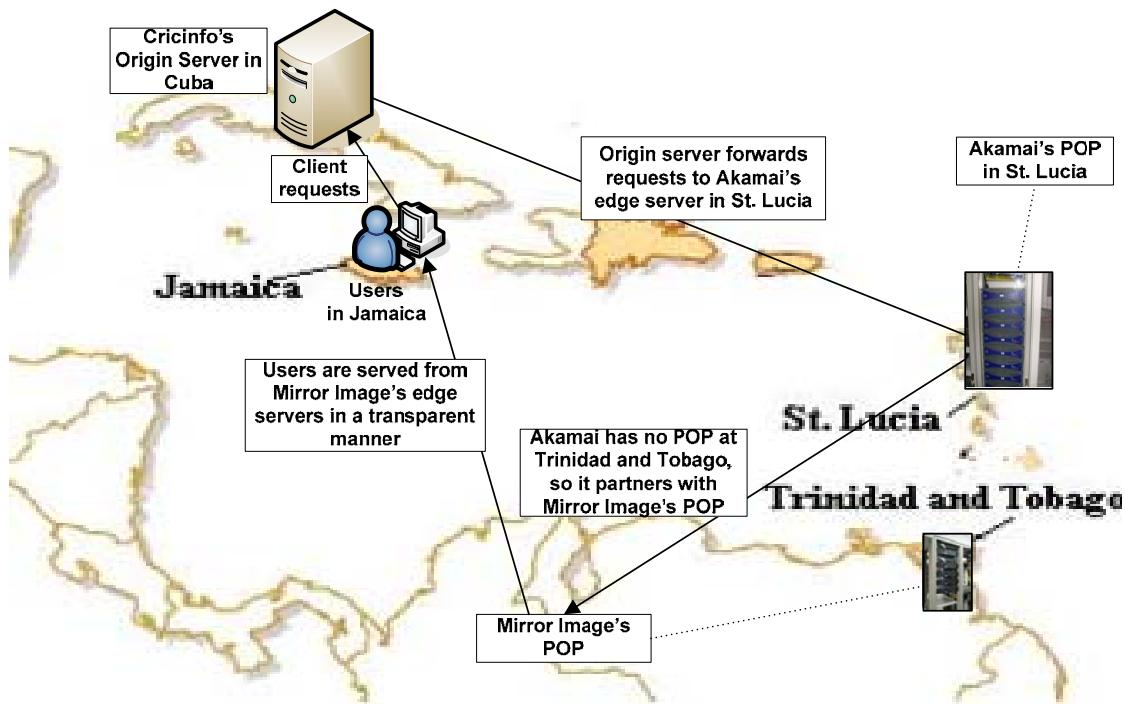

**Figure 2:** CDN internetworking scenario

To make content services an Internet infrastructure service, vendors have implemented content service networks (CSN) [5], which act as another network infrastructure layer built upon CDNs and provide next generation of CDN services. CSN appears to be another variation of the conventional CDN as it is just another layer of network infrastructure built around CDN. This logical separation between content and services under the 'Content Delivery/Distribution' and 'Content Services' domain, is undesirable considering the on-going trend in content networking. Hence, a unified content network which supports the coordinated composition and delivery of content and services, is highly desirable.

## 3. A Model for Peering CDNs

In this section, we present the model of an open, scalable, and Service Oriented Architecture (SOA) based system. This system assists the creation of open CSDNs (Content and Service Delivery Networks) that scale and support sharing of resources with other CSDNs through cooperation and coordination. Thus, it helps to overcome the problem of creation of *islands* of CDNs, to ensure the quality of services based on SLA negotiation, and to find a solution to the problem of the logical separation between Content Delivery Network (CDN) and Content Services Network (CSN). We propose a Virtual Organization (VO) [2] model for forming CSDNs that not only support sharing of Web servers within their own networks, but also with other CSDNs. The architecture of such a system is shown in Figure 3.

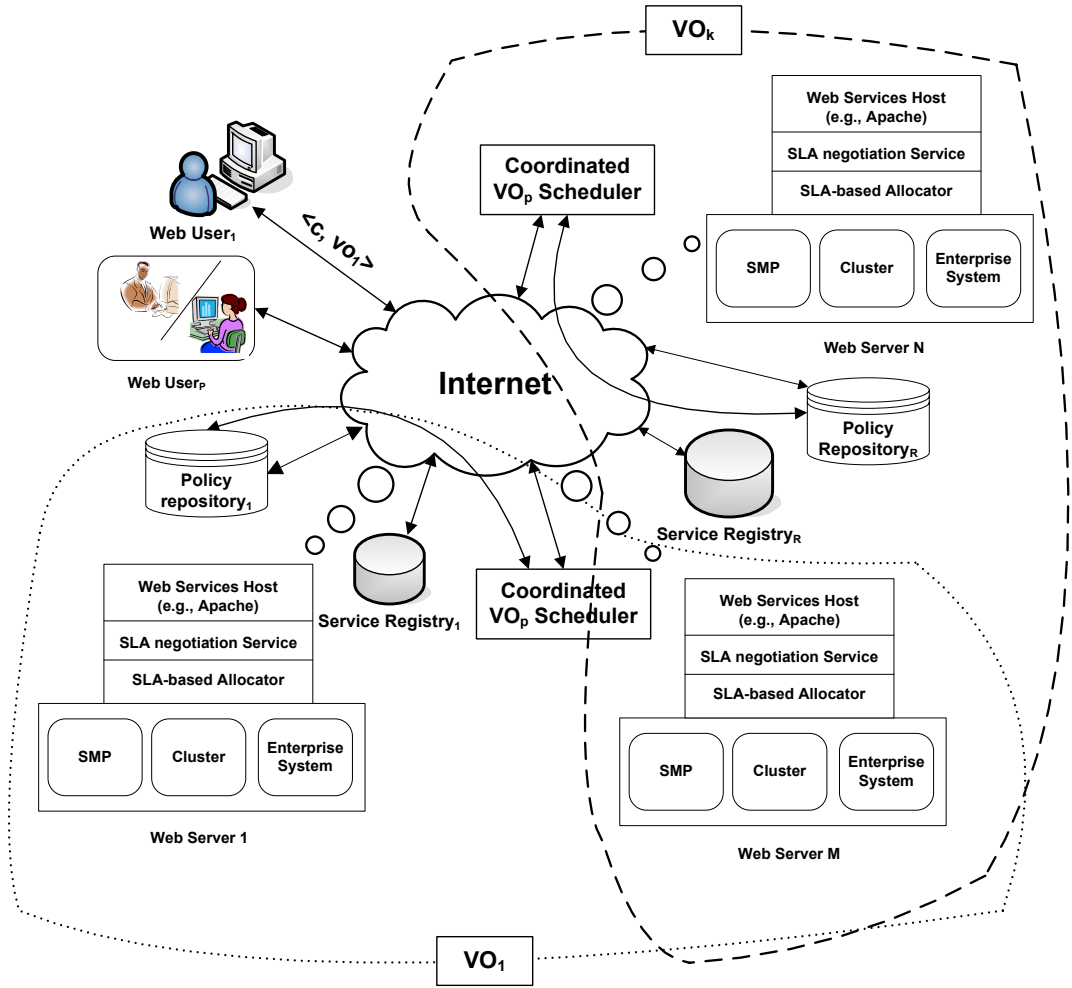

**Figure 3:** Architecture of open, scalable, Service-Oriented Architecture (SOA) based system to assist the creation of cooperative and coordinated CSDNs (Content and Service Delivery Networks)

## 4. Components of CSDN

Some of the elements of the CSDN architecture presented above are described below:

**Web Server:** This entity is the most important element of a Content and Service Delivery Network (CSDN). A CSDN is formed consisting of the Web servers using the VO [2] model. Web servers are responsible for storing contents and value-added services as infrastructure services, and delivering them in a reliable and cooperative manner. Web servers within each CSDN are capable of delivering contents and services in order to meet QoS requirements of end-users (i.e. Web users). The structure of a Web server can be divided into two layers: an *overlay layer* and the *core*. In the *overlay layer*, a Web server consists of a Web service host (e.g. Apache and Tomcat), an SLA-negotiation service module and an SLA-based Allocator. The negotiation module with the help of coordinated VO scheduler is responsible for cooperation and coordination with other Web servers (located in a local or global CSDN) through SLA-based negotiation. The SLA-based allocator is put in place in order to deliver contents and services based on the negotiated SLAs with other Web servers of local or global CSDNs. The *core* of the Web server consists of high performance computing systems such as SMPs (Symmetric Multiprocessors), Cluster and/or other enterprise systems (e.g. desktop grids). The underlying devices and tools of the Web server are responsible for storing content and services, and assist in responding client requests in a timely and reliable manner in order to meet the negotiated QoS requirements. For content and service location and routing, the underlying technologies of the Web servers perform on-demand cooperative caching through coordination with other web servers. Efficiently balancing the load across different Web servers is critical to produce the required QoS. Hence, Web servers are adopted with appropriate load and resource distribution strategies.

**Web user:** Web users are the clients who request for content and/or services from the Web servers. The requested content is served to the clients either from the Web server receiving client requests or from any other Web server within the VO, which is *closest* to the Web user.

**Coordinated VO scheduler:** A coordinated VO scheduler is put in each VO which is responsible for ensuring collaboration and coordination with other CSDNs though policy exchange and scheduling of contents and services.

**Service registry:** A service registry enables VOs to register their cluster resources. SLA negotiator Service and Allocator module use this service registry to negotiate QoS parameters and resource allocation to maximize the potential of cooperative CSDNs.

**Policy repository:** A policy repository is used to store the policies generated by the administrators. These policies are a set of rules to administer, manage, and control access to VO resources. They provide a way to consistently mange the components deploying complex technologies.

## 5. Summary and Future Directions

In this article, we present an open, scalable and Service-Oriented Architecture (SOA) based system to assist the creation of open Content and Service Delivery Networks (CSDNs). Thus we address the issue of content networks scaling and support sharing of resources with other CSDNs. We propose a VO model for forming CSDNs and a policy framework within the VO model. It will support management and sharing of contents and services not only within their own networks, but also with other CSDNs. Delivery of resources (i.e., contents and services) in such an environment will meet QoS requirements of end-users according to the negotiated SLA. The realization of such a system is expected to be a timely contribution to the ongoing trend of Content Networking. Work is in progress on peering CDNs with joint collaboration between the GRIDS laboratory, University of Melbourne and the DSN laboratory, RMIT University, Australia.

## Resources